\documentclass[prb,twocolumn,showpacs]{revtex4}

\usepackage{graphicx}
\usepackage{dcolumn}
\usepackage{bm}
\usepackage[dvips]{color}
\usepackage{longtable}

\def\mrm{\mathrm}

\def\etal{{\it et al. }}
\def\goto{\rightarrow}

\def\mrm{\mathrm}

\def\goto{\rightarrow}

\def\xih{\hat{\xi}}

\def\deli0{\delta_{\sigma_i 0}}
\def\delj0{\delta_{\sigma_j 0}}

\def\t{|t|}
\def\smax{\langle s_\mrm{max} \rangle}

\definecolor{darkgreen}{rgb}{0.0, 0.4, 0.0}

\def\n+{_{n+1}}

\begin{document}


\preprint{}

\title{
Criticality governed by the stable renormalization fixed point 
\\
of the Ising model in the hierarchical small-world network
}

\author{Tomoaki Nogawa}
\email{nogawa@serow.t.u-tokyo.ac.jp}
\affiliation{%
Department of Mathematics, Tohoku University, 
6-3-09, Aramaki-Aza-Aoba, Sendai, Miyagi 980-8579, Japan
}
\author{Takehisa Hasegawa}
\affiliation{%
Graduate School of Information Science, 
Tohoku University, 
6-3-09, Aramaki-Aza-Aoba, Sendai, Miyagi 980-8579, Japan
}%
\author{Koji Nemoto}
\affiliation{%
Department of Physics, Hokkaido University,
Kita 10 Nisi 8, Kita-ku
Sapporo, Hokkaido 060-0810, Japan
}%

\begin{abstract}
We study the Ising model in a hierarchical small-world network 
by renormalization group analysis, and 
find a phase transition between an ordered phase and a critical phase, 
which is driven by the coupling strength of the shortcut edges. 
Unlike ordinary phase transitions, which are related 
to unstable renormalization fixed points (FPs), 
the singularity in the ordered phase of the present model is governed by the FP 
that coincides with the stable FP of the ordered phase. 
The weak stability of the FP yields peculiar criticalities including logarithmic behavior. 
On the other hand, the critical phase is related to a nontrivial FP, 
which depends on the coupling strength and is continuously connected to the ordered FP 
at the transition point.
We show that this continuity indicates the existence of a finite correlation-length-like quantity 
inside the critical phase, which diverges upon approaching the transition point. 
\end{abstract}


\maketitle



Recently, various physical phenomena in non-Euclidean graphs 
have been studied especially in the context of complex networks, 
and their properties have been found to be beyond the scope of the conventional theory 
for Euclidean graphs \cite{Dorogovtsev08}.
Of particular interest are systems regarded as infinite dimensional in the sense 
that the equidistant surface $S_r$ of radius $r$ grows exponentially as 
$S_r \propto e^{y_d r}$ with a positive constant $y_d$, 
which is faster than any power function $r^{d-1}$ as in $d$-dimensional Euclidean graphs.
Typical examples are trees and hyperbolic lattices \cite{Shima06}.
Remarkably, such infinite-dimensional systems often exhibit the critical phases 
in which the (nonlinear) susceptibility diverges 
\cite{Hinczewski06, Berker09, Nogawa-Hasegawa09}. 
Although the critical phase is also observed in some Euclidean systems, 
e.g., the quasi-long-range ordered phase in the two-dimensional XY model 
\cite{Berezinskii72, Kosterlitz73},
the critical phases in infinite dimensional systems are considered 
to be due to rather geometrical effects. 
Indeed, the exponential growth of a graph admits the divergence of the susceptibility $\chi$, 
which is calculated by the integral of the two-point correlation function
\footnote{
We had better to consider local susceptibility 
when the system is inhomogeneous (nontransitive) \cite{Bauer05}.
}, 
even with finite correlation length $\xih$ \cite{Nogawa-Hasegawa09b};
for $\xih > 1/y_d$ 
\begin{equation}
\chi \propto \int_0^{\ln N/y_d} dr S_r e^{-r/\xih} \propto  N^\psi, 
\quad 
\psi \equiv 1 - 1/y_d \xih
\label{eq:chi-cr}
\end{equation}
diverges in the thermodynamic limit $N \to \infty$. 
Here we set the upper bound $\ln N/y_d$ to ensure $\chi \propto N$ for $\xih=\infty$. 
A critical phase, if it exists, lies between a disordered phase with $\xih < 1/y_d$ 
and an ordered phase with $\xih = \infty$. 
Such a phase with a fractal exponent $0<\psi<1$ is actually observed 
in the percolation transitions in enhanced trees \cite{Nogawa-Hasegawa09}, 
hyperbolic lattices \cite{Baek09}, hierarchical graphs \cite{Hasegawa10, Boettcher12}, 
and growing random networks \cite{Hasegawa10c}.

\begin{figure}[t]
\begin{center}
\includegraphics[trim=  102 610 100 -200, scale=0.395, clip, angle=0]{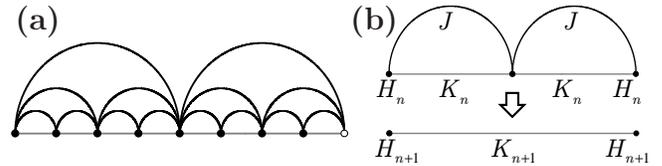}
\\
\vspace{-2.25cm}
\hspace{-3.4cm} {\bf {\large (a) \hspace{3.7cm}(b) }}
\\
\vspace{1.95cm}
\end{center}
\vspace{-5mm}
\caption{\label{fig:graph}
(a) Hierarchical small-world network with eight nodes.
The periodic boundary condition is imposed on the horizontal direction. 
(b) One-step decimation of the central spin in the partial sum of partition function. 
This is an inverse procedure to grow the graph in (a).
}
\end{figure}

The property of the phase transitions between a critical phase 
and an ordered phase is an interesting issue. 
Quite recently, some models in the simple hierarchical network 
shown in Fig.~\ref{fig:graph}(a) were investigated to examine these transitions.
This network is very useful because rigorous real-space renormalization 
is possible for various models in the simplest way. 
Furthermore various types of phase transition are observed depending on the model used, 
e.g., a discontinuous transition of the bond percolation model \cite{Boettcher12}, 
equivalent to the 1-state Potts model \cite{Kasteleyn69, Fortuin71}, 
and continuous transitions with a power-law singularity (PLS) or an essential singularity (ES) 
for the $q$-state Potts model with $q \ge 3$ \cite{Nogawa-Hasegawa12}. 
These are observed in other graphs 
\cite{Krapivsky04, Hinczewski06, Boettcher09, Berker09, Nogawa-Hasegawa09}.
Thus this hierarchical network is a good stage to investigate 
what determines the type of phase transitions in a systematic way.
In particular the 2-state Potts model, equivalent to the Ising model, 
stands between $q=1$ and $q=3$, and has special importance to understand 
how the transition class changes.

In this Rapid Communication, we study the phase transition of the 2-state Potts model 
in the network mentioned above by renormalization group (RG) analysis, 
which reveals a new class of phase transition governed by the stable fixed point.  
Furthermore the singularity for $q=2$ is found to be special 
due to the marginal bifurcation of the RG fixed point (FP) 
between pitchfork type for $q<2$ and saddle-node type for $q>2$.


We consider the 2-state Potts model in the hierarchical small-world network 
shown in Fig.~\ref{fig:graph}(a). 
This consists of one-dimensional backbone edges  (BBEs) 
and nested shortcut edges (SCEs). 
The spin variable $\sigma$, taking a value 0 or 1, is put on every node of the network. 
The dimensionless energy function is written as 
\begin{eqnarray}
-\frac{E}{k_\mrm{B} T} 
= \!\! \sum_\mrm{\langle i,j \rangle \in BBE} \!\! K \delta_{\sigma_i \sigma_j} 
+ \!\! \sum_\mrm{\langle i,j \rangle \in SCE} \!\! J \delta_{\sigma_i \sigma_j}
+ \sum_i H \delta_{\sigma_i 0}, 
\end{eqnarray}
where $K$ and $J$ are the coupling constants on BBEs and SSEs, respectively, 
and $H(>0)$ denotes the external magnetic field coupled with the state 0.


Let us consider the partial sum of the partition function 
$Z=\prod_i \sum_{\sigma_i} e^{-E/k_BT}$ 
over the states of spins in the youngest generation having two neighbors 
[see Fig.~\ref{fig:graph}(b)]. 
This is rigorously equivalent to replacing the parameters as 
\begin{eqnarray}
C_{n+1} e^{
K_{n+1} \delta_{\sigma_1 \sigma_2} 
+ (H_{n+1}/2)( \delta_{\sigma_1 0} + \delta_{\sigma_2 0} )
}
\nonumber \\
= C_n^2 e^{
(H_{n}/2)( \delta_{\sigma_1 0} + \delta_{\sigma_2 0} )
}
\sum_\sigma e^{
(K_n+J)( \delta_{\sigma \sigma_1}  + \delta_{\sigma \sigma_2} )
+ H_n \delta_{\sigma 0} 
},
\end{eqnarray}
with the recursion relations:
\begin{eqnarray}
g_{n+1} &\!\! = \!\!& g_n + 2^{-(n+1)} \! \left[ K_n + J -  A^{(1)}_n/2 + A^{(2)}_n/2 + A^{(3)}_n \right],
\nonumber
\label{eq:RG4f}
\\
K_{n+1} &\!\! = \!\!& - (K_n + J) + A^{(1)}_n/2 + A^{(2)}_n/2 - A^{(3)}_n,
\label{eq:RG4K}
\\
H_{n+1} &\!\! = \!\!& H_n + A^{(1)}_n - A^{(2)}_n,
\label{eq:RG4H}
\end{eqnarray}
where $g_n = 2^{-n} \ln C_n$, 
$e^{A^{(1)}_n} = e^{2 (K_n+J) + H_n} + 1$, 
$e^{A^{(2)}_n} = e^{2 (K_n+J)} + e^{H_n}$, and 
$e^{A^{(3)}_n} = e^{H_n} + 1$.

\begin{figure}[t]
\begin{center}
\includegraphics[trim=-1 50 -1 20,scale=0.30,clip]{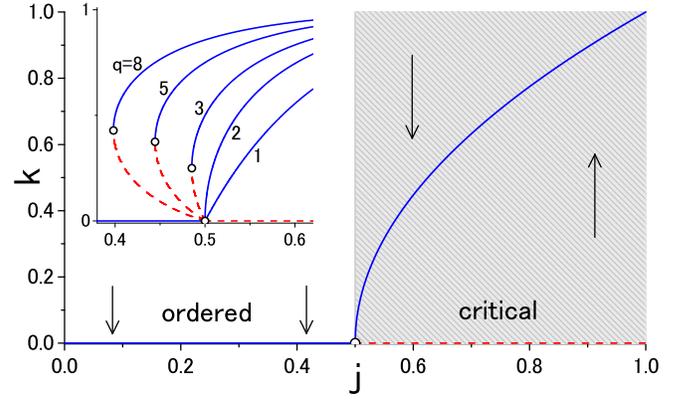}
\end{center}
\vspace{-5mm}
\caption{\label{fig:PD}
(color online) 
Phase diagram in the $j$ vs $k$ plain. 
The shaded region indicates the critical phase.
The (blue) solid line and (red) dashed line denote 
stable and unstable fixed lines, respectively. 
The arrows indicate the RG flow of $k$. 
The inset shows the fixed lines for $q$=1, 2, 3, 5, and 8.
}
\end{figure}

For $H=0$, the recursion relation is rewritten as 
\begin{eqnarray}
k_{n+1} - k_n = - \frac{k_n \left[ j^2 k_n^2 - (2j-1) \right]}{1 + j^2 k_n^2}, 
\label{eq:k-diff}
\end{eqnarray}
where we put $k_n = e^{-K_n}$ and $j = e^{-J}$.
The FP is given by $k_{n+1}=k_n=k^*$, which is solved as 
\begin{eqnarray}
k^* = 0, \pm \sqrt{2t/(1+t)}, 
\quad
t \equiv 2j-1.
\end{eqnarray}
There is a pitchfork bifurcation point (PF\,BP) at $(j, k)=(1/2,0)$; 
the stable FP is $k^*=0$ for $j \le 1/2$ and $k^*=\sqrt{2t/(1+t)}$ for $j>1/2$
(see Fig.~\ref{fig:PD}). 
The region where the flow goes to the nontrivial fixed {\it line} is regarded as a critical phase 
because each FP on the line represents a self-similar structure 
like an ordinary critical point. 
It is noteworthy that in this renormalization procedure 
the couplings on the SCEs remain the same 
while the ones on the BBEs are updated. 
The SCE couplings $0< J \ll 1$ prevent $K_n$ from converging to zero,
and therefore the critical phase appears
instead of a disordered phase.
A similar mechanism is observed in the RG analysis of the Ising models 
exhibiting the critical phases, e.g., in graphs: the decorated (2,2)-flower \cite{Hinczewski06} 
and the Hanoi network with average degree 5 (HN5) \cite{Boettcher11}.

When $j$ increases with $k$ fixed, a transition 
from the ordered phase to the critical phase occurs at $j=j_c=1/2$, 
irrespectively of the value of $k$. 
Remarkably, the FP at $j_c$ corresponds to $K=\infty (k=0)$ as well as in the ordered phase 
$j<j_c$ and is stable, so that the critical behavior is quite unlike the conventional one 
governed by unstable FP (US FP). 
This yields a curious singularity, as shown below.



Equation~(\ref{eq:k-diff}) is approximated by a differential equation: 
\begin{eqnarray}
\frac{d s_n}{d n} = -\frac{2 s_n (s_n + \t)}{1+s_n}, 
\quad
s_n \equiv k_n^2 j^2.
\label{eq:dsdn}
\end{eqnarray}
This is solved as 
$ 
e^{-2n \t} = \frac{s_n}{s_0} \left( \frac{s_0 + \t}{s_n + \t} \right)^{1-\t}
$ 
leading to 
\begin{eqnarray}
s_n \approx \frac{\t}{e^{2 n \t}-1}=\left\{ 
\begin{array}{ccc}
1/2n & \mrm{for} & n \t \ll 1
\\
\t e^{-2 n \t} & \mrm{for} & n \t \gg 1
\end{array}
\right. 
\label{eq:flow_sol}
\end{eqnarray}
for $\t \ll s_0$. 
Since $s_n$ goes to zero for $n \goto \infty$, 
it is (and thus $K$ is) an irrelevant parameter.


In the lowest order of $H_n$, Eq.~(\ref{eq:flow_sol}) is still valid 
as a solution of Eq.~(\ref{eq:RG4K}).
On the other hand, Eq.~(\ref{eq:RG4H}) is rewritten as 
\begin{eqnarray}
H_{n+1} = H_n + \ln \frac{ e^{2(K_n+J) + H_n} + 1}{ e^{2(K_n+J)} + e^{H_n}}
= \frac{2 H_n}{ 1 + s_n} + O(H_n^2).
\label{eq:rg_H_n}
\end{eqnarray}
Note that we leave $s_n$ in the first term; 
the cross term of $s_n$ and $H_n$ remains 
and thus this is beyond the linear stability analysis. 
By putting $H_n = 2^n e^{-x_n}$, we obtain 
\begin{eqnarray}
x_n 
\approx \sum_{\ell=n_0}^{n-1} s_\ell 
\approx \int_{n_0}^{n} \frac{\t d \ell}{e^{2 \ell \t}-1} 
= \frac{1}{2} \ln \frac{1-e^{-2n \t}}{1-e^{-2 n_0 \t}},
\end{eqnarray}
where we substitute Eq.~(\ref{eq:flow_sol}) into the summation 
and approximate it by the integral.
Here we give the lower bound of summation $n_0$ 
to avoid the divergence of $s_\ell$ at $\ell=0$. 
Finally we obtain 
\begin{eqnarray}
H_n = H e^{n y_d} u_n
\nonumber \\ 
u_n \approx \sqrt{ \frac{\t}{ 1 - e^{-2n\t} } }
\propto \left \{
\begin{array}{ccc}
n^{-1/2} & \mrm{for} & n \t \ll 1
\\
\t^{1/2} & \mrm{for} & n \t \gg 1
\end{array}
\right. \! ,
\label{eq:H_scl}
\end{eqnarray}
where we use $y_d=\ln2$. 
Note that the factor $u_n$ does not appear in the linear-order treatment 
for $s_n$ and $H_n$.



The RG evolution Eq.~(\ref{eq:H_scl}) leads to a scale-invariant formula of the free energy as 
\begin{eqnarray}
g(H,N^{-1}) = e^{-n y_d} g(H u_n e^{n y_d}, N^{-1} e^{n y_d}) 
\label{eq:free_energy}
\end{eqnarray}
for $n \gg 1$. 
Here $t$ is not included explicitly in the arguments of $g$ 
because it is (and thus $J$ is) not a scaling field but just an external parameter. 
The irrelevant scaling field $s_n$ is also omitted. 
However the effect of $s_n$ is indirectly included in $H_n$ as in Eq.~(\ref{eq:rg_H_n}) 
and gives the $t$ dependence as in Eq.~(\ref{eq:H_scl}).
The first and second order derivatives of $g$ with $H$ are written as 
\begin{eqnarray}
g_H(H,N^{-1}) &=& u_n g_H(H u_n e^{n y_d}, N^{-1} e^{n y_d}),  
\label{eq:g_H}
\\
g_{H^2}(H,N^{-1}) &=& u_n^2 e^{n y_d}
g_{H^2}(H u_n e^{n y_d}, N^{-1} e^{n y_d}), 
\label{eq:g_HH}
\end{eqnarray}
respectively.
In the following, we show the behaviors of the magnetization $m=g_H$ 
and the susceptibility $\chi=g_{H^2}$ in three asymptotic regimes: 
(i)~$t<0$, $H=0$ and $N=\infty$, 
(ii)~$t=0$, $H>0$ and $N=\infty$, 
and (iii)~$t=0$, $H=0$ and $N<\infty$.

(i)~For $t < 0$, $H=0$ and $N=e^{ny_d} \goto \infty$, we obtain 
\begin{eqnarray}
m &=& \t^{1/2} g_H(0,1) \propto \t^{1/2}, 
\label{eq:m-t}
\\
\chi &=& N \t g_{H^2}(0,1) \propto N \t , 
\label{eq:chi-t}
\end{eqnarray}
with $u_n \goto \t^{1/2}$. 
The magnetization shows the PLS as if it were an ordinary second order transition. 
Within the linear analysis $u_n=1$, 
$m$ does not vanish at $t=0$, and the transition would look discontinuous. 
Surprisingly, $\chi$ diverges in the whole ordered phase, 
meaning the coexistence of a divergent fluctuation and a nonzero order parameter. 
Such coexistence may be realized by spatial segregation into  
the fluctuating region and the ordered region around the root node, 
the left most node in Fig.~\ref{fig:graph}(a).

(ii) For $t \goto 0$ and $N=e^{ny_d}$, $u_n \goto n^{-1/2}$, 
Eqs.~(\ref{eq:g_H}) and (\ref{eq:g_HH}) become 
\begin{eqnarray}
g_H(H,N^{-1}) &=& n^{-1/2} g_H(H n^{-1/2} e^{n y_d}, 1), 
\label{eq:g_H_t0}
\\
g_{H^2}(H,N^{-1}) &=& n^{-1} e^{n y_d} 
g_{H^2}(H n^{-1/2} e^{n y_d}, 1).
\label{eq:g_HH_t0}
\end{eqnarray}
We set $n$ so as to satisfy $H n^{-1/2} e^{n y_d}=1$, 
which is approximately solved as $e^{n y_d} \approx  \ln (H^{-1})/y_d H$.
By substituting this into Eqs.~(\ref{eq:g_H_t0}) and (\ref{eq:g_HH_t0}), we obtain
\begin{eqnarray}
m &\approx& \sqrt{ 
\frac{y_d}{\ln(H^{-1})} } g_H(1, 1)
\propto \left[ \ln(H^{-1}) \right]^{-1/2},
\label{eq:m-H}
\\
\chi &\approx& \frac{y_d}{\ln H^{-1}}
\frac{\ln{H^{-1}}}{y_d H} g_{H^2}(1,1) 
\propto \frac{1}{H}.
\label{eq:chi-H}
\end{eqnarray}

(iii) For $t=0$ and $H=0$, Eqs.~(\ref{eq:g_H_t0}) and (\ref{eq:g_HH_t0}) become 
\begin{eqnarray}
m &=& \sqrt{ \frac{y_d}{\ln N} } g_H(0, 1)
\propto \left( \ln N  \right)^{-1/2}, 
\\
\chi &=& \frac{N}{\ln N} g_{H^2}(0, 1) \propto \frac{N}{\ln N}.
\label{eq:chi-N}
\end{eqnarray}

\begin{figure}[t]
\begin{center}
\includegraphics[trim=40 30 140 20,scale=0.30,clip]{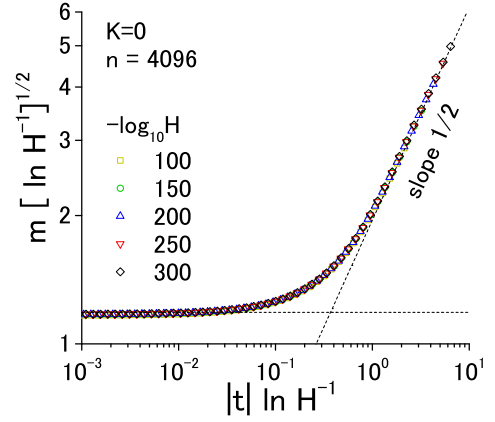}
\end{center}
\vspace{-5mm}
\caption{\label{fig:m-t}
(color online) 
Scaling plot of the magnetization in the ordered phase.
Data with $m<0.20$ are used. 
}
\end{figure}

On the crossover between the two limits, Eqs.~(\ref{eq:m-t}) and (\ref{eq:m-H}), 
we expect a one-parameter scaling formula for $N=\infty$ such as 
\begin{equation}
m = t^{1/2} \tilde{m} ( t \ln H^{-1}) 
= \left[ \ln(H^{-1}) \right]^{-1/2} \tilde{\tilde{m}} ( t \ln H^{-1}), 
\end{equation}
which is confirmed in Fig.~\ref{fig:m-t}. 
We numerically calculate $m$ by integrating the recursion equations 
of $g_n$, $K_n$, $H_n$, and their derivatives.





Next, we investigate the property of the critical phase ($t>0$).  
By considering the linear stability of Eqs.~(\ref{eq:RG4K}) and (\ref{eq:RG4H}) 
at the points on the fixed line, we obtain  
$\tilde{K}_{n+1} = e^{n y_K(t)} \tilde{K}_n$ 
and $H_{n+1} = e^{n y_H(t)} H_{n}$ 
where $\tilde{K}_n \equiv K_{n} - K^*(t)$, 
$ 
y_K(t) = \ln \frac{1-t}{1+t},  
\quad \mrm{and} \quad 
y_H(t) = \ln \frac{2}{1+t}. 
$ 
These lead to $\tilde{K}_n \propto e^{n y_K}$ and $H_n \propto e^{n y_H}$.
While $y_K(t)$ is negative and therefore $\tilde{K}$ is irrelevant in the critical phase, 
$y_H(t)$ is positive and approaches $y_d=\ln2$ from below as $t \goto 0_+$.

\begin{figure}[b]
\begin{center}
\includegraphics[trim=40 55 80 30,scale=0.320,clip]{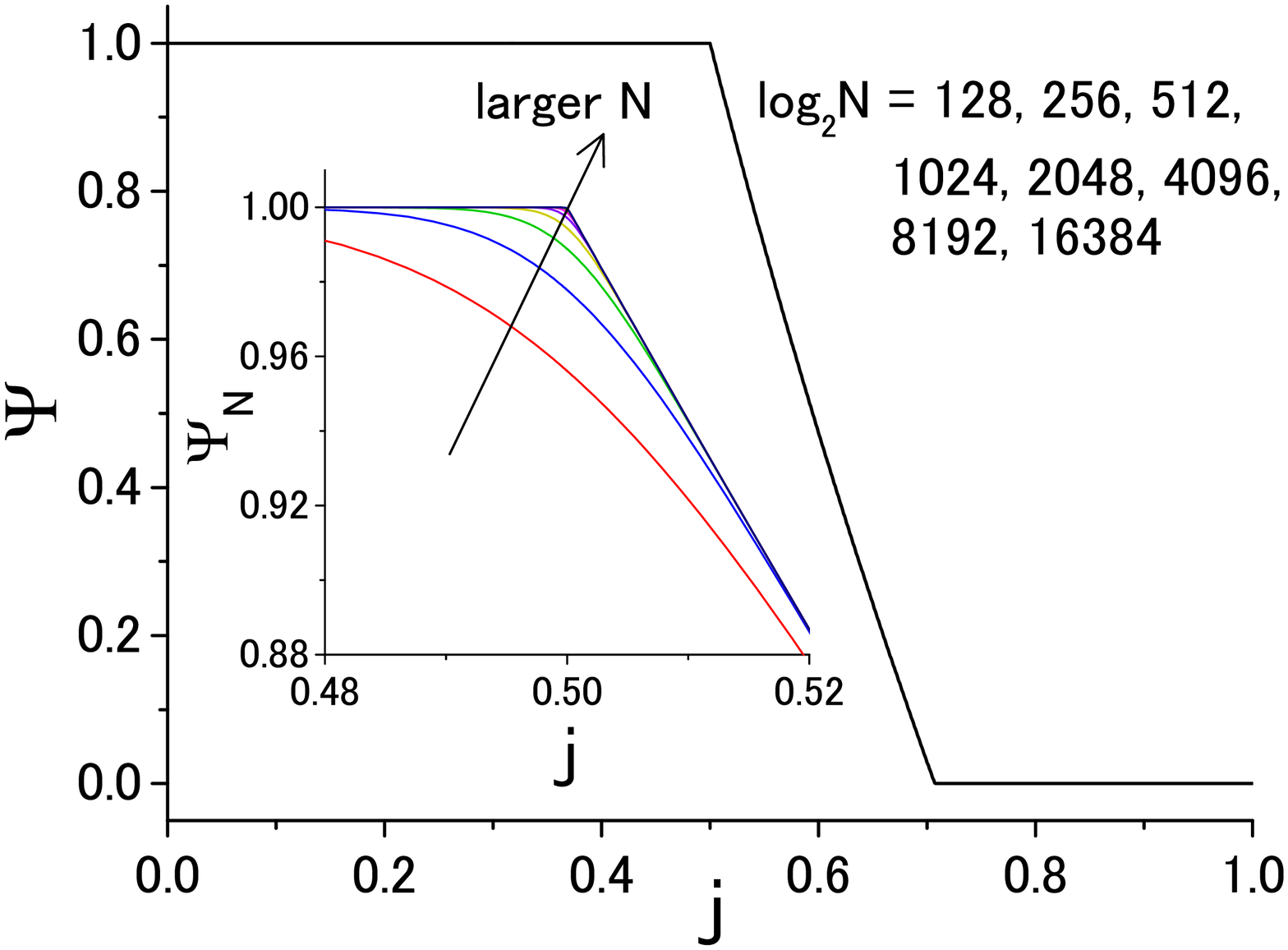}
\end{center}
\vspace{-5mm}
\caption{\label{fig:psi-j}
(color online) 
The fractal exponent is plotted with respect to the BBE coupling constant. 
The inset enlarges the finite size behavior $\psi_N$ around $j=j_c=1/2$. 
}
\end{figure}

The free energy in the critical phase satisfies 
\begin{eqnarray}
g(H,N^{-1}) = e^{-n y_d} 
g(H e^{n y_H(t)}, N^{-1}e^{n y_d} ).
\nonumber
\end{eqnarray}
Again we omit the irrelevant parameters $\tilde{K}$ and $t$ from the arguments, 
but $t$-dependence is included in $y_H$. 
The $\ell$-th order derivative of $g$ is written as 
\begin{eqnarray}
g_{H^\ell}(H,N^{-1}) = e^{n(\ell y_H(t) - y_d)} 
g_{H^\ell}(H e^{n y_H(t)}, N^{-1}e^{n y_d} ). 
\nonumber
\end{eqnarray}
Since $y_H(t) < y_d$, the first order derivative, i.e., the magnetization, 
is zero for $H=0$ and $N=e^{n y_d} \goto \infty$ in the critical phase. 
The susceptibility for $H=0$ and $N=e^{n y_d}$ is 
\begin{eqnarray}
\chi = N^{\psi(t)} g_{H^2}(0,1) \propto N^{\psi(t)}
\\ 
\psi(t) = 2 \frac{y_H(t)}{y_d} - 1 = 1 - \frac{2}{\ln 2} \ln (1+t). 
\label{eq:psi-t}
\end{eqnarray}
For $1/2<j<1/\sqrt{2}$, 
$\psi$ varies from 1 to 0 as shown in Fig.~\ref{fig:psi-j}. 
In this region, $\chi$ diverges in the thermodynamic limit.
For $j>1/\sqrt{2}$, $\chi$ is finite 
but higher order derivatives, i.e., nonlinear susceptibilities, diverge; 
the $\ell$th order derivative of the free energy with $H$ diverges 
in the region where $\ell y_H(t) > y_d$.



\begin{figure}[t]
\begin{center}
\includegraphics[trim=20 20 140 25,scale=0.30,clip]{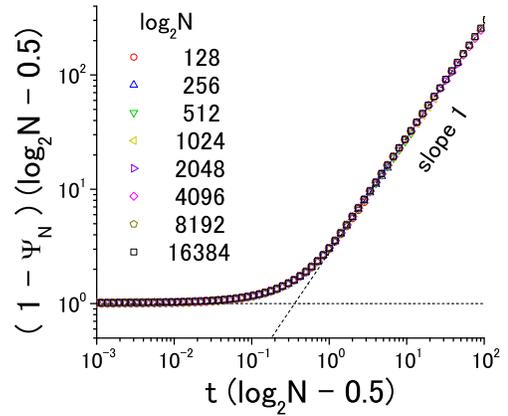}
\end{center}
\vspace{-5mm}
\caption{\label{fig:psi-t_scl}
(color online) 
Scaling plot of the fractal exponent $\psi_N$ 
in the critical phase. 
Data with $\psi_N>0.60$ are used. 
}
\end{figure}

As noted in Eq.~(\ref{eq:chi-cr}), $\psi$ is related to the correlation length $\xih$, 
and the fact that $\psi$ approaches 1 from below means the divergence of $\xih$. 
Here we consider the finite size effect of this singularity.
The inset of the Fig.~\ref{fig:psi-j} shows the fractal exponent for size $N$ 
defined as $\psi_N \equiv \log_2 [\chi(N)/\chi(N/2)]$.
If we assume a power-law divergence of $\xih$, 
a finite size scaling formula, 
\begin{eqnarray}
\xih^{-1} & \equiv & 1 - \psi_N = t^{\nu} F(t^{\nu} [\ln N]^\mu) 
\nonumber \\
& = & \left\{
\begin{array}{ccc}
t^{\nu} & \mrm{for} & t^\nu [\ln N]^\mu  \gg 1
\\
\left[ \ln N \right]^{-\mu} & \mrm{for} & t^\nu [\ln N]^\mu \ll 1
\end{array}
\right. ,
\label{eq:scaling_L}
\end{eqnarray}
is expected for $H=0$.
If $\ln N$ is a proper cutoff length, $\mu$ should be unity.
Equations (\ref{eq:psi-t}) and (\ref{eq:chi-N}) lead to 
$1 - \psi_N \propto t$
and 
$1 - \psi_N \propto [ \ln N ]^{-1}$, respectively. 
Thus $\nu = 1$ and $\mu = 1$.
We confirm the scaling behavior Eq.~(\ref{eq:scaling_L}) 
in Fig.~\ref{fig:psi-t_scl}.


We have investigated the 2-state Potts (Ising) model in the simple hierarchical network 
with the real-space RG method. 
The phase transition between the ordered and critical phases 
is governed by the PF\,BP at $K=\infty$ 
independently of the bare value of $K$. 
The singular behavior of the present model is summarized in Table~\ref{tbl:summary}, 
where we also show the Potts model with $q=1$ \cite{Boettcher12} 
and $q \ge 3$ \cite{Nogawa-Hasegawa12}. 
See also the inset of Fig.~\ref{fig:PD} for the fixed lines for various $q$ systems.

Boettcher \etal studied the bond percolation model, i.e., the 1-state Potts model, 
in the same graph without backbone ($K=0$) \cite{Boettcher12}, 
and found a PF bifurcation of the RG\,FP similar to the present model ($q=2$).
They also calculate the maximum cluster size $\smax$, 
which is a quantity corresponding to unconnected susceptibility $\tilde{\chi} \equiv \chi + N m^2$ 
because both quantities are defined as a summation of the two-point correlation. 
For $q=1$, however, discontinuity of $\smax/N$ is observed at the transition point. 
This is in contrast to the fact that  $\tilde{\chi}/N \propto t$ for $q=2$.
By taking the analytic continuation of the recursion relation, 
corresponding to Eq.~(\ref{eq:k-diff}), 
for $q \ge 3$ in Ref.~\cite{Nogawa-Hasegawa12}, 
we obtain the PF\,BP $(j,k)=(1/2,0)$ for $q \le 2$ and 
the recursion relation at $j=1/2$: 
$k_{n+1} - k_{n} = - \frac{2-q}{4}k^2 - \frac{q-1}{4} k^3 + O(k^4)$. 
Thus the FP of $q=2$ is marginal and has weaker stability. 
This is presumably the origin of the continuity of the transition.

For $q \ge 3$, the saddle-node (SN) bifurcation of the FP is observed \cite{Nogawa-Hasegawa12}. 
Consequently, two kinds of singularity appear depending on $k=e^{-K}$: 
ES corresponding to SN\,BP for $k \ge k_\mrm{SN}$ 
and PLS corresponding to USFP for $k < k_\mrm{SN}$. 
The SN\,BP $(j_\mrm{SN},  k_\mrm{SN})$ approaches the PF\,BP $(1/2, 0)$ as $q \goto 2$. 
While the PLS and ES for $q \ge 3$ are governed by USFP 
and marginally unstable SN\,BP, respectively, 
the phase transition for $q \le 2$ is governed by the stable FP. 
Thus the generalized scaling theory in Ref.~\cite{Nogawa-Hasegawa12} assuming instability of a FP 
cannot be applied to the latter. 
We emphasize that the stability or instability of the FP of the transition point 
is the most fundamental criterion of phase transitions.

\begin{table}[b]
\begin{tabular}{ccccc}
\tabcolsep=18cm
$q$ & FP & $m$ & $\chi$ & $\hat{\xi} \equiv 1/(1-\psi)$ \\
\hline
$1$ & \ PF\,BP & \ -- & $N |t|^0$,\ -- & $t^{-2},\ [\ln N]^2$
\\
$2$ & \ PF\,BP & \ $\sqrt{|t|},\ [\ln H^{-1}]^{-1/2}$ & $N |t|,\ H^{-1}$ & $t^{-1}, \ln N$
\\
$\ge3$ & \ SN\,BP & \ $e^{-|t|^{-1/2}},\ H^{1/\delta}$ 
& $e^{|t|^{-1/2}},\ H^{1/\delta-1}$ & $t^0$
\\
$\ge3$ & \ USFP & \ $|t|^\beta, H^{1/\delta}$ & $|t|^{-\gamma},\ H^{1/\delta-1}$ & $t^0$ 
\end{tabular}
\caption{
Summary of singularities.
We show the singular formula of $m$ and $\chi$ in the ordered phase $t<0$ 
and of $\hat{\xi}$ in the critical phase $t>0$. 
A discontinuous change is noted as $t^0$.
The fact that $\hat{\xi} \propto [\ln N]^2$ for $q=1$ is our preliminary result 
of a Monte Carlo simulation. 
}
\label{tbl:summary}
\end{table}


\begin{thebibliography}{19}
\expandafter\ifx\csname natexlab\endcsname\relax\def\natexlab#1{#1}\fi
\expandafter\ifx\csname bibnamefont\endcsname\relax
  \def\bibnamefont#1{#1}\fi
\expandafter\ifx\csname bibfnamefont\endcsname\relax
  \def\bibfnamefont#1{#1}\fi
\expandafter\ifx\csname citenamefont\endcsname\relax
  \def\citenamefont#1{#1}\fi
\expandafter\ifx\csname url\endcsname\relax
  \def\url#1{\texttt{#1}}\fi
\expandafter\ifx\csname urlprefix\endcsname\relax\def\urlprefix{URL }\fi
\providecommand{\bibinfo}[2]{#2}
\providecommand{\eprint}[2][]{\url{#2}}

\bibitem[{\citenamefont{Dorogovtsev et~al.}(2008)\citenamefont{Dorogovtsev,
  Goltsev, and Mendes}}]{Dorogovtsev08}
\bibinfo{author}{\bibfnamefont{S.~N.} \bibnamefont{Dorogovtsev}},
  \bibinfo{author}{\bibfnamefont{A.~V.} \bibnamefont{Goltsev}},
  \bibnamefont{and} \bibinfo{author}{\bibfnamefont{J.~F.~F.}
  \bibnamefont{Mendes}}, \bibinfo{journal}{Rev. Mod. Phys.}
  \textbf{\bibinfo{volume}{80}}, \bibinfo{pages}{1275} (\bibinfo{year}{2008}).

\bibitem[{\citenamefont{Shima and Sakaniwa}(2006)}]{Shima06}
\bibinfo{author}{\bibfnamefont{H.}~\bibnamefont{Shima}} \bibnamefont{and}
  \bibinfo{author}{\bibfnamefont{Y.}~\bibnamefont{Sakaniwa}},
  \bibinfo{journal}{J. Stat. Mech.} p. \bibinfo{pages}{08017}
  (\bibinfo{year}{2006}).

\bibitem[{\citenamefont{Hinczewski and Berker}(2006)}]{Hinczewski06}
\bibinfo{author}{\bibfnamefont{M.}~\bibnamefont{Hinczewski}} \bibnamefont{and}
  \bibinfo{author}{\bibfnamefont{A.~N.} \bibnamefont{Berker}},
  \bibinfo{journal}{Phys. Rev. E} \textbf{\bibinfo{volume}{73}},
  \bibinfo{pages}{066126} (\bibinfo{year}{2006}).

\bibitem[{\citenamefont{Berker et~al.}(2009)\citenamefont{Berker, Hinczewski,
  and Netz}}]{Berker09}
\bibinfo{author}{\bibfnamefont{A.~N.} \bibnamefont{Berker}},
  \bibinfo{author}{\bibfnamefont{M.}~\bibnamefont{Hinczewski}},
  \bibnamefont{and} \bibinfo{author}{\bibfnamefont{R.~R.} \bibnamefont{Netz}},
  \bibinfo{journal}{Phys. Rev. E} \textbf{\bibinfo{volume}{80}},
  \bibinfo{pages}{041118} (\bibinfo{year}{2009}).

\bibitem[{\citenamefont{Nogawa and
  Hasegawa}(2009{\natexlab{a}})}]{Nogawa-Hasegawa09}
\bibinfo{author}{\bibfnamefont{T.}~\bibnamefont{Nogawa}} \bibnamefont{and}
  \bibinfo{author}{\bibfnamefont{T.}~\bibnamefont{Hasegawa}},
  \bibinfo{journal}{J. Phys. A: Math. Theor.} \textbf{\bibinfo{volume}{42}},
  \bibinfo{pages}{145001} (\bibinfo{year}{2009}{\natexlab{a}}).

\bibitem[{\citenamefont{Berezinskii}(1972)}]{Berezinskii72}
\bibinfo{author}{\bibfnamefont{V.~L.} \bibnamefont{Berezinskii}},
  \bibinfo{journal}{Sov. Phys. JETP} \textbf{\bibinfo{volume}{34}},
  \bibinfo{pages}{610} (\bibinfo{year}{1972}).

\bibitem[{\citenamefont{Kosterlitz and Thouless}(1973)}]{Kosterlitz73}
\bibinfo{author}{\bibfnamefont{J.~M.} \bibnamefont{Kosterlitz}}
  \bibnamefont{and} \bibinfo{author}{\bibfnamefont{D.~J.}
  \bibnamefont{Thouless}}, \bibinfo{journal}{J. Phys. C}
  \textbf{\bibinfo{volume}{6}}, \bibinfo{pages}{1181} (\bibinfo{year}{1973}).

\bibitem[{\citenamefont{Nogawa and
  Hasegawa}(2009{\natexlab{b}})}]{Nogawa-Hasegawa09b}
\bibinfo{author}{\bibfnamefont{T.}~\bibnamefont{Nogawa}} \bibnamefont{and}
  \bibinfo{author}{\bibfnamefont{T.}~\bibnamefont{Hasegawa}},
  \bibinfo{journal}{J. Phys. A: Math. Theor.} \textbf{\bibinfo{volume}{42}},
  \bibinfo{pages}{478002} (\bibinfo{year}{2009}{\natexlab{b}}).

\bibitem[{\citenamefont{Baek et~al.}(2009)\citenamefont{Baek, Minnhagen, and
  Kim}}]{Baek09}
\bibinfo{author}{\bibfnamefont{S.~K.} \bibnamefont{Baek}},
  \bibinfo{author}{\bibfnamefont{P.}~\bibnamefont{Minnhagen}},
  \bibnamefont{and} \bibinfo{author}{\bibfnamefont{B.~J.} \bibnamefont{Kim}},
  \bibinfo{journal}{Phys. Rev. E} \textbf{\bibinfo{volume}{79}},
  \bibinfo{pages}{011124} (\bibinfo{year}{2009}).

\bibitem[{\citenamefont{Hasegawa et~al.}(2010)\citenamefont{Hasegawa, Sato, and
  Nemoto}}]{Hasegawa10}
\bibinfo{author}{\bibfnamefont{T.}~\bibnamefont{Hasegawa}},
  \bibinfo{author}{\bibfnamefont{M.}~\bibnamefont{Sato}}, \bibnamefont{and}
  \bibinfo{author}{\bibfnamefont{K.}~\bibnamefont{Nemoto}},
  \bibinfo{journal}{Phy. Rev. E} \textbf{\bibinfo{volume}{82}},
  \bibinfo{pages}{046101} (\bibinfo{year}{2010}).

\bibitem[{\citenamefont{Boettcher et~al.}(2012)\citenamefont{Boettcher, Singh,
  and Ziff}}]{Boettcher12}
\bibinfo{author}{\bibfnamefont{S.}~\bibnamefont{Boettcher}},
  \bibinfo{author}{\bibfnamefont{V.}~\bibnamefont{Singh}}, \bibnamefont{and}
  \bibinfo{author}{\bibfnamefont{R.~M.} \bibnamefont{Ziff}},
  \bibinfo{journal}{Nature Communications} \textbf{\bibinfo{volume}{3}},
  \bibinfo{pages}{787} (\bibinfo{year}{2012}).

\bibitem[{\citenamefont{Hasegawa and Nemoto}(2010)}]{Hasegawa10c}
\bibinfo{author}{\bibfnamefont{T.}~\bibnamefont{Hasegawa}} \bibnamefont{and}
  \bibinfo{author}{\bibfnamefont{K.}~\bibnamefont{Nemoto}},
  \bibinfo{journal}{Phy. Rev. E} \textbf{\bibinfo{volume}{81}},
  \bibinfo{pages}{051105} (\bibinfo{year}{2010}).

\bibitem[{\citenamefont{Kasteleyn and Fortuin}(1969)}]{Kasteleyn69}
\bibinfo{author}{\bibfnamefont{P.~W.} \bibnamefont{Kasteleyn}}
  \bibnamefont{and} \bibinfo{author}{\bibfnamefont{C.~M.}
  \bibnamefont{Fortuin}}, \bibinfo{journal}{J. Phys. Soc. Jap.}
  \textbf{\bibinfo{volume}{26}}, \bibinfo{pages}{11} (\bibinfo{year}{1969}).

\bibitem[{\citenamefont{Fortuin and Kasteleyn}(1972)}]{Fortuin71}
\bibinfo{author}{\bibfnamefont{C.~M.} \bibnamefont{Fortuin}} \bibnamefont{and}
  \bibinfo{author}{\bibfnamefont{P.~W.} \bibnamefont{Kasteleyn}},
  \bibinfo{journal}{Physica} \textbf{\bibinfo{volume}{57}},
  \bibinfo{pages}{536} (\bibinfo{year}{1972}).

\bibitem[{\citenamefont{Nogawa et~al.}(2012)\citenamefont{Nogawa, Hasegawa, and
  Nemoto}}]{Nogawa-Hasegawa12}
\bibinfo{author}{\bibfnamefont{T.}~\bibnamefont{Nogawa}},
  \bibinfo{author}{\bibfnamefont{T.}~\bibnamefont{Hasegawa}}, \bibnamefont{and}
  \bibinfo{author}{\bibfnamefont{K.}~\bibnamefont{Nemoto}},
  \bibinfo{journal}{Phys. Rev. Lett.} \textbf{\bibinfo{volume}{108}},
  \bibinfo{pages}{255703} (\bibinfo{year}{2012}).

\bibitem[{\citenamefont{Krapivsky and Derrida}(2004)}]{Krapivsky04}
\bibinfo{author}{\bibfnamefont{P.~L.} \bibnamefont{Krapivsky}}
  \bibnamefont{and} \bibinfo{author}{\bibfnamefont{B.}~\bibnamefont{Derrida}},
  \bibinfo{journal}{Physica A} \textbf{\bibinfo{volume}{340}},
  \bibinfo{pages}{714} (\bibinfo{year}{2004}).

\bibitem[{\citenamefont{Boettcher et~al.}(2009)\citenamefont{Boettcher, Cook,
  and Ziff}}]{Boettcher09}
\bibinfo{author}{\bibfnamefont{S.}~\bibnamefont{Boettcher}},
  \bibinfo{author}{\bibfnamefont{J.~L.} \bibnamefont{Cook}}, \bibnamefont{and}
  \bibinfo{author}{\bibfnamefont{R.~M.} \bibnamefont{Ziff}},
  \bibinfo{journal}{Phys. Rev. E} \textbf{\bibinfo{volume}{80}},
  \bibinfo{pages}{041115} (\bibinfo{year}{2009}).

\bibitem[{\citenamefont{Boettcher and Brunson}(2011)}]{Boettcher11}
\bibinfo{author}{\bibfnamefont{S.}~\bibnamefont{Boettcher}} \bibnamefont{and}
  \bibinfo{author}{\bibfnamefont{C.~T.} \bibnamefont{Brunson}},
  \bibinfo{journal}{Phys. Rev. E} \textbf{\bibinfo{volume}{83}},
  \bibinfo{pages}{021103} (\bibinfo{year}{2011}).

\bibitem[{\citenamefont{Bauer et~al.}(2005)\citenamefont{Bauer, Coulomb, and
  Dorogovtsev}}]{Bauer05}
\bibinfo{author}{\bibfnamefont{M.}~\bibnamefont{Bauer}},
  \bibinfo{author}{\bibfnamefont{S.}~\bibnamefont{Coulomb}}, \bibnamefont{and}
  \bibinfo{author}{\bibfnamefont{S.~N.} \bibnamefont{Dorogovtsev}},
  \bibinfo{journal}{Phys. Rev. Lett.} \textbf{\bibinfo{volume}{94}},
  \bibinfo{pages}{200602} (\bibinfo{year}{2005}).

\end{thebibliography}



\end{document}